\documentstyle[epsf,12pt]{article}
  \def\lsim{\raise0.3ex\hbox{$<$\kern-0.75em\raise-1.1ex\hbox{$\sim$}}}
\def\gsim{\raise0.3ex\hbox{$>$\kern-0.75em\raise-1.1ex\hbox{$\sim$}}}
\def\noi{\noindent}

\def\nn{\nonumber}
\def\bea{\begin{eqnarray}}  \def\eea{\end{eqnarray}}
\def\beq{\begin{equation}}   \def\eeq{\end{equation}}
\def\sq{\hbox {\rlap{$\sqcap$}$\sqcup$}}
\def\beeq{\begin{eqnarray}} \def\eeeq{\end{eqnarray}}
\newcommand\mysection{\setcounter{equation}{0}\section}
\renewcommand{\theequation}{\thesection.\arabic{equation}}
\newcounter{hran} \renewcommand{\thehran}{\thesection.\arabic{hran}}
\def\bmini{\setcounter{hran}{\value{equation}}
  \refstepcounter{hran}\setcounter{equation}{0}
  \renewcommand{\theequation}{\thehran\alph{equation}}\begin{eqnarray}}
\def\bminiG#1{\setcounter{hran}{\value{equation}}
\refstepcounter{hran}\setcounter{equation}{-1}
\renewcommand{\theequation}{\thehran\alph{equation}}
\refstepcounter{equation}\label{#1}\begin{eqnarray}}
\def\emini{\end{eqnarray}\relax\setcounter{equation}{\value{hran}}
\renewcommand{\theequation}{\thesection.\arabic{equation}}}

\begin{document}
\begin{center}
\vbox to 1 truecm {}
{\large \bf Confinement, Monopoles and Wilsonian} \\
{\large \bf  Effective Action} \\ 

\vskip 8 truemm
{\bf Ulrich Ellwanger}\footnote{e-mail : ellwange@qcd.th.u-psud.fr}\\
Laboratoire de Physique Th\'eorique et Hautes
Energies\footnote{Laboratoire associ\'e 
au Centre National de la Recherche Scientifique - URA D0063} \\
Universit\'e de Paris XI, B\^atiment 210,
F-91405 Orsay Cedex, France \\
\end{center}

\vskip 1 truecm

\begin{abstract}
An effective low energy action for Yang-Mills theories is proposed,
which invokes an additional 
auxiliary field $H_{\mu \nu}$ for the field strength $F_{\mu \nu}$. For
a particular relation 
between the parameters of this action a gluon propagator with a $1/p^4$ 
behaviour for $p^2 \to 0$ in the Landau gauge is
obtained. The abelian subsector of this action admits a duality
transformation, where the dual 
action contains a Goldstone boson $\varphi$ as the dual of $H_{\mu
\nu}$, and corresponds to an 
abelian Higgs model in the broken phase describing the condensation of
magnetic charges. The 
Wilsonian renormalization group equations for the parameters of the
original action are integrated 
in some approximation, and we find that the relation among the
parameters associated with confinement 
appears as an infrared attractive fixed point.  \end{abstract}

\vskip 1 truecm

\noi LPTHE Orsay 97-44 \par
\noi September 1997 \par
\newpage
\pagestyle{plain}
\mysection{Introduction} 
\hspace*{\parindent}
Presently we are still lacking a proper field theoretical formulation,
let alone a quantitative
description, of confinement in continuum QCD. An illustrative picture is
provided by the 
Mandelstam-t'Hooft dual superconductor mechanism of confinement
\cite{1r}. Here it is assumed that 
monopoles (with respect to the $U(1)$ subgroups of $SU(3)_{color}$)
condense in the QCD vacuum. Then 
it is supposed that chromo-electric flux tubes confine chromo-electric
charges in the same way as 
magnetic flux tubes confine magnetic charges inside a superconductor,
where electric charges have condensed. \par

It is very difficult, however, to formulate this idea in the context of
a local quantum field 
theory, and even more to prove, that this is a consequence of the
infrared dynamics of QCD. There 
are indications for condensed monopoles in the QCD vacuum from lattice
QCD \cite{2r}, but this does 
not help to find a description of monopole condensation in the
continuum. \par 

Monopoles are known to arise as classical solutions (solitons) in field
theoretic models like the 
Georgi-Glashow model ($SO(3)$ Yang-Mills theory with Higgs scalars in
the adjoint representation) 
\cite{3r}. t'Hooft has pointed out that similar structures could also
appear in pure Yang-Mills 
theories, since the role of the fundamental Higgs scalars could
eventually be played by some 
composite fields \cite{1r,3r}. In any case the monopoles can be
understood, in the framework of such 
models, as defects in space-time of $U(1)$ gauge fields which arise once
the unitary gauge is chosen 
\cite{1r,3r,4r,5r}. \par

It is notoriously difficult, however, to describe such defects (or
solitons in general) in terms of 
quantum fields \cite{6r} and hence to study quantities like effective
potentials. Moreover, even if 
one assumes the presence of fields with magnetic charges, a formulation
of a quantum field theory is 
far from trivial: a Lagrangian for $U(1)$ gauge fields, in the
presence of both electrically 
and magnetically charged fields, is necessarily non-Poincar\'e-invariant
(a fixed vector $n_{\mu}$ 
has to be introduced as in axial gauges) and either non-local or
requires the introduction of a 
second gauge potential \cite{7r,4r}. \par

These latter problems can be circumvented by the introduction of dual
fields: in the case of a 
scalar field $\varphi$ with magnetic charge, e.g., the dual field would
be an antisymmetric tensor 
$H_{\mu \nu}$ (in $d = 4$). In terms of $H_{\mu \nu}$ it is possible to
formulate Lagrangians, 
which are both Poincar\'e invariant and local \cite{8r,9r}. \par

Employing, on the other hand, dual gauge potentials, it is at least
possible to formulate low 
energy models for QCD, which realize the Mandelstam-t'Hooft dual
superconductor mechanism of 
confinement \cite{10r,11r}. Clearly, these models are entirely based on
assumptions and do not allow 
to make contact with the bare Lagrangian of QCD. \par

The purpose of the present paper is twofold: first, we propose a low
energy effective action for 
$SU(N)$ Yang-Mills theories, which invokes an antisymmetric tensor field
$H_{\mu \nu}^a$ in the 
adjoint representation of the gauge group. We show that, for suitably
chosen parameters, this 
effective action implies confinement in the sense of a ${1 / p^4}$
infrared singularity of the 
gluon propagator in the Landau gauge. The abelian subsector of the
action admits a duality 
transformation, and the corresponding dual action is the one of an
abelian Higgs model with a 
Goldstone boson in the broken phase. Since accordingly the dual electric
charge has condensed in the 
vacuum, it follows that the original magnetic charge has condensed in
the vacuum. \par 

The role of the field $H_{\mu \nu}^a$ in this action is the one of a
composite field for the field 
strength tensor $F_{\mu \nu}^a$. This can be made explicit by
interpreting the low e\-ner\-gy 
effective action as a Wilsonian effective action, which satisfies
Wilsonian exact renormalization 
group equations (ERGEs). The ERGEs allow to relate the low e\-ner\-gy
effective action in a continuous 
manner to a bare Lagrangian or high energy action, in which the field
$H_{\mu \nu}^a$ appears only 
without space-time derivatives and can be eliminated by its algebraic
equations of motion of the form 
$H_{\mu \nu}^a = F_{\mu \nu}^a$. \par

The second purpose of the paper is thus to set up the system of ERGEs
for the effective action, and 
to solve them within some approximation, which allows to keep track of
all the important parameters. 
We find indeed that the ERGE flow leads us from the bare Yang-Mills
action at high scales (including 
the auxiliary field $H_{\mu \nu}^a$) towards a low energy effective
action, whose parameters are 
such that confinement in the above sense (${1 / p^4}$ behavior of the
gluon propagator and 
monopole condensation) occurs. We have thus available both a model for a
low energy effective action, 
which realizes the t'Hooft-Mandelstam mechanism of confinement, and a
formalism, which allows us to 
compute the corresponding parameters from the bare Lagrangian. \par

Let us, to close the introduction, repeat the essential ideas behind the
Wilsonian ERGE approach 
\cite{12r}: the starting point is a partition function, in which an
infrared cutoff $k$ for all 
propagators is introduced. From the partition function one obtains the 
 generating functionals ${\cal G}_k(J)$ and its Legendre 
transform $\Gamma_k(\varphi)$, which are the usual generating
functionals of connected and 
one-particle-irreducible Green functions respectively, where, however,
only internal propagators 
with $p^2 > k^2$ contribute. For both functionals ${\cal G}_k(J)$ and
$\Gamma_k(\varphi )$ one can 
write down exact but simple differential equations with respect to $k$,
the ERGEs. For very large 
values of $k$, the functionals become simply related to the bare
Lagrangian of the corresponding 
theory, the knowledge of which can thus serve as boundary condition for
the integration of the 
ERGEs at some large value $k = \Lambda$. Integrating the ERGEs down to
$k = 0$ provides us with the 
full physical functionals as the effective action $\Gamma_0(\varphi )$.
With the integration 
starting at some large but finite value $k = \Lambda$, $\Gamma_0(\varphi
)$ corresponds to 
the effective action in the presence of an UV cutoff $\Lambda$, and is
obtained as a function of the 
finite parameters of the corresponding bare Lagrangian. Actually, in
order to study the 
renormalizability of a theory, the $\Lambda$ dependence of
$\Gamma_0(\varphi )$ can also be studied 
using ERGEs (which now correspond to exact differential equations with
respect to $\Lambda$ 
\cite{13r}). In the recent years, much progress has been made in
employing this approach both for 
scalar theories and gauge theories \cite{14r} -- \cite{19r}. \par

A particularly useful aspect of this method is the fact that it can be 
extended in a
straightforward way to functionals of local composite operators 
\cite{17r,18r}: these can be
introduced in any theory by introducing sources for the composite 
operators in the partition
function. After a Legendre transformation one obtains an effective action 
which depends on both
fundamental and composite fields. This effective action can also be studied 
using ERGEs. Its
dependence on the composite field(s), at the starting scale $k = \Lambda$ of 
the ERGE flow, has
to be trivial: the action $\Gamma_{\Lambda}$ depends on the composite field 
just through a
quadratic term without derivatives, which would allow - in principle - for 
its trivial
elimination by its algebraic equation of motion. The dependence of the full 
physical effective
action $\Gamma_0$ at $k = 0$ on the composite field depends on the dynamics 
of the theory and is
obtained as before by integration of the ERGEs. If the full propagator of the 
composite field, as
obtained from the quadratic terms in $\Gamma_0$, possesses a pole in $p^2$ 
with the correct sign
of the residuum, the composite field corresponds to a propagating bound 
state. Even if such a
pole is absent, however, the presence of the composite (auxiliary) field 
can be helpful for
simplifying the description of the dynamics of the theory. \par

Below we will introduce a composite field $H_{\mu \nu}^a$ for the field 
strength operator $F_{\mu
\nu}^a(A)$ in non-abelian Yang-Mills theories. Although we do not expect 
$H_{\mu \nu}^a$ to
correspond to a physical propagating bound state in such theories, we 
will find that its presence
can be very helpful for simplifying the description of confinement. \par

In the next section we will discuss several general properties of an 
effective action invoking a
composite field $H_{\mu \nu}^a$: its formal definition, and the conditions 
under which one
obtains confinement. We will describe the duality transformation (of the 
abelian subsector),
which allows to interpret confinement as the condensation of monopoles. \par

In section 3 we will study the ERGE flow for the effective action in some appro\-xi\-ma\-tion and find
that it interpolates indeed between the bare Yang-Mills action at the 
starting scale $k =
\Lambda$ and a confining action at low scales. Section 4 is devoted to 
conclusions and an outlook.   

\mysection{Effective Action for Yang-Mills Theories with an Antisymmetric 
Tensor Field} 
\hspace*{\parindent}
Let us start this section with the definitions of the generating 
functionals in Euclidean Yang-Mills
theory in the presence of an infrared cutoff $k$, and a source 
$K_{\mu \nu}^a$ coupled to the
composite operator 

\beq
\label{2.1}
F_{\mu \nu}^a = \partial_{\mu} \ A_{\nu}^a - \partial_{\nu} \ A_{\mu}^a 
+ g \ f_{bc}^a \ A_{\mu}^b \
A_{\nu}^c \quad . \eeq

\noi We will use the short-hand notations
 
\bea
&&J\cdot A \equiv \int {\cal D}_p \ J_{\mu}^a (-p) \ A_{\mu}^a(p) 
\ \ \hbox{etc.} \quad ; \nn \\
&& {\cal D}_p \equiv {d^4p \over (2 \pi )^4} \quad .
 \label{2.2} 
\eea

\noi Including the standard covariant gauge fixing, ghosts and the 
infrared 
cutoff, the expression for the generating functional of connected Green 
functions ${\cal G}_k(J, \chi, \bar{\chi}, K)$ is 

\bea
\exp \left ( - {\cal G}_k (J, \chi, \bar{\chi}, K) \right ) &=& \int 
{\cal D}_{reg} (A, c , \bar{c}) \exp \Big (
- (S_{YM} + S_g + \Delta S_k)  \nn \\
\label{2.3}
&& + J \cdot A + \bar{\chi} \cdot c + \chi \cdot \bar{c} +
K\cdot F + K \cdot K \Big ) \quad . \eea

\noi Here $S_{YM}$ is the standard Yang-Mills action

\beq
\label{2.4}
S_{YM} = {1 \over 4} \int d^4x \ F_{\mu \nu}^a \ F_{\mu \nu}^a \quad ,
\eeq

\noi and $S_g$ the gauge fixing and ghost part:

\beq
\label{2.5}
S_g = \int d^4x \left [ {1 \over 2 \alpha} \partial_{\mu} \ A_{\mu}^a 
\ \partial_{\nu} \ A_{\nu}^a +
\partial_{\mu} \ \bar{c}^a \left ( \delta_{ac} \ \partial_{\mu} + g \ 
f^a_{bc} \ A_{\mu}^b \right )
c^c \right ] \quad . \eeq

\noi The term $\Delta S_k$ implements the infrared cutoff for the gauge 
and ghost fields; it is
given by

\beq
\label{2.6}
\Delta S_k = \int {\cal D}_p \left [ {1 \over 2} A_{\mu}^a(-p) \ 
R_{\mu \nu}^k(p^2) \ A_{\nu}^a(p) +
\bar{c}^a(-p) \ R_g^k(p^2) \ c^a(p) \right ] \quad . \eeq

\noi The functions $R_{\mu \nu}^k$ and $R_g^k$ modify the gauge and 
ghost propagators such that
modes with $p^2 \ll k^2$ are suppressed. Convenient choices are 

\bea
\label{2.7}
&&R_{\mu \nu}^k(p^2) = \left ( p^2 \ \delta_{\mu \nu} + \left 
( {1 \over \alpha} - 1 \right )
p_{\mu} \ p_{\nu} \right ) {e^{-p^2/k^2} \over 1 - e^{-p^2/k^2}} 
\quad ,  \nn \\ 
&&R_g^k(p^2) = p^2{e^{-p^2/k^2} \over 1 - e^{-p^2/k^2}} \quad .
\eea
 
\noi The functions $R_{\mu \nu}^k(p^2)$ and $R_g^k(p^2)$ vanish 
for $k^2 = 0$, and are finite for
$p^2 \ll k^2$ such that the full gluon and ghost propagators are 
infrared finite in this regime. \par

$K$ denotes the source for the composite operator $F_{\mu \nu}^a$, 
and we have used the freedom
\cite{20r} to add a term quadratic in $K$, which simplifies the boundary 
conditions for the
Wilsonian generating functionals ${\cal G}_k$ and $\Gamma_k$ for $k 
\to \infty$ \cite{18r}. \par

The index ``reg'' attached to the path integral measure indicates an 
ultraviolet regularization,
which is required to make the path integral well defined. Finally the 
generating
functionals will be defined entirely through integrated ERGEs with 
suitable boundary conditions,
which corresponds implicitely to a re\-gu\-la\-ri\-za\-tion involving 
higher derivatives. Its explicit
form is, however, not required here. \par

The underlying gauge symmetry constrains the physical generating 
functionals ${\cal G}_0$ and
$\Gamma_0$ severely via Slavnov-Taylor identities. These imply 
modified Slavnov-Taylor
identities for the functionals ${\cal G}_k$ and $\Gamma_k$ in the 
presence of a non-vanishing
infrared cutoff $k$ \cite{15r,16r}. In principle their proper formulation 
requires the introduction
of sources coupled to the BRST variations of all the fields in the path 
integral eq.~(\ref{2.3}).
Although we will also make use of these identities (e.g. in order to 
relate the three-gluon,
four-gluon and ghost-gluon vertices) we have, for simplicity, not 
shown these sources explicitely
in eq.~(\ref{2.3}). In particular they are not required for the 
ERGEs themselves. \par

The effective action in the presence of the infrared cutoff $k$ is 
defined, as usual, through the Legendre transform 

\beq
\label{2.8}
\widetilde{\Gamma}_k(A, c, \bar{c}, H) = {\cal G}_k (J, \chi, 
\bar{\chi}, K) + 
J \cdot A + \bar{\chi} \cdot c + \chi
\cdot \bar{c} + K \cdot H \quad . \eeq 

\noi Sometimes it is more convenient to subtract the cutoff terms 
$\Delta S_k$ from $\widetilde{\Gamma}_k$ and to work with

\beq
\label{2.9}
\Gamma_k (A, c, \bar{c}, H) = \widetilde{\Gamma}_k(A, c, \bar{c}, H) - 
\Delta S_k \quad .
\eeq

\noi From the path integral (\ref{2.3}) and the Legendre transform 
(\ref{2.8}) it is straighforward
to derive the ERGE for the cutoff effective action \cite{14r}-\cite{19r}:

\beq
\label{2.10}
\partial_k \Gamma_k = {1 \over 2} \int {d^4p \over (2 \pi )^4}  
\partial_k \ R^k(p^2)_{ij}
  \left ( {\delta^2 \widetilde{\Gamma}_k \over \delta \ 
\bar{\varphi}_{\ell}(-p) \ \delta
\varphi_m(p)} \right )^{-1}_{ji} \quad . \eeq

\noi Here the fields $\varphi_{i} \equiv (A_{\mu}^a , c^a, \bar{c}^a, 
H_{\mu \nu}^a)$ denote all
possible field appearing as arguments of $\Gamma_k$ or 
$\widetilde{\Gamma}_k$, the index $i$
corresponding to the field type and the Lorentz and gauge group indices. 
\par

The matrix $R_{ij}^k$ has non-vanishing matrix elements only in the 
subsectors $(A_{\mu}^a,
A_{\nu}^a)$ and $(\bar{c}^a, c^a)$, cf. the form of $\Delta S_k$ given 
in eq.~(\ref{2.6}). The inverse functional \par

\noi $(\delta^2 \widetilde{\Gamma}_k/\delta \bar{\varphi}_{\ell} \delta 
\varphi_m)^{-1}_{ji}$, however, has to be
constructed on the complete space spanned by $(A_{\mu}, c, \bar{c}, 
H_{\mu \nu})$ including the
auxiliary field $H_{\mu \nu}^a$. \par

It is clear that, in general, $\Gamma_k$ contains terms with arbitrary 
powers of the fields, and
with arbitrary powers of derivatives or momenta $p$. Moreover, the 
modified Slavnov-Taylor
identities \cite{15r,16r} require, in general, the presence of all 
terms, which are invariant under
the global part of the gauge group (and have vanishing ghost number). 
Let us, to start with, write
down explicitely the BRST invariant terms in $\Gamma_k$ which a) are 
quadratic in the field strength
$F_{\mu \nu}^a$, the auxiliary field $H_{\mu \nu}^a$ or the ghosts, b) 
contain the lowest
non-trivial number of (covariant) derivatives: 

\bea
\label{2.11}
\Gamma_k &=& {Z \over 4} \left ( F_{\mu \nu} \right )^2 - {n \over 2} 
F_{\mu \nu} \ H_{\mu \nu} + {m^2
\over 4} \left ( H_{\mu \nu} \right )^2 \nn \\
&+& {h \over 2} \left ( D_{\mu} \ \widetilde{H}_{\mu \nu} \right )^2 + 
{\beta \over 2} \left (
D_{\mu} \ H_{\mu \nu} \right )^2 + Z_g \ \partial_{\mu} \ \bar{c} \ 
D_{\mu} \ c \nn \\ 
&+& {1 \over 2 \alpha} \left ( \partial_{\mu} \ A_{\mu} \right )^2 + 
\Delta S_k + \cdots 
 \eea

\noi Here $\widetilde{H}_{\mu \nu}^a$ is defined by $\widetilde{H}_{\mu 
\nu}^a = {1 \over 2}
\varepsilon_{\mu \nu \rho \sigma} \ H_{\rho \sigma}^a$, and the covariant 
derivative $D_{\mu}$,
acting on fields $\varphi^a$ in the adjoint representation of the gauge 
group, by

\beq
\label{2.12}
D_{\mu} \ \varphi^a = \partial_{\mu} \ \varphi^a + \bar{g} \ f_{bc}^a \ 
A_{\mu}^b \ \varphi^c \quad . 
\eeq 

\noi The dots in (\ref{2.11}) denote terms of higher order in the fields, 
the field strength or
derivatives, and terms which are not BRST invariant (like a gluon mass term) 
but fixed in terms of
the other ones through the modified Slanov-Taylor identities. \par

All parameters $Z$, $n$, $m$, $h$, $\beta$, $Z_g$ and $\alpha$ appearing 
in (\ref{2.11}) depend on the scale $k$, which is equal to the infrared
cutoff in $\Delta S_k$ introduced in the path integral (\ref{2.3}) and 
present also $\Gamma_k$
itself. The corresponding ERGEs can be obtained by expanding both sides 
of eq.~(\ref{2.10}) to
second order in the corresponding fields, and to the corresponding order 
of derivatives. These ERGEs
will be derived and studied in section 3. \par

The boundary conditions for the integration of the ERGEs are imposed at 
some large scale $k =
\Lambda$, where we require $\Gamma_{k=\Lambda}$ to correspond to the bare 
Lagrangian of the theory
up to additional terms required by the modified Slavnov-Taylor identities. 
Implicitely this leads to
a physical effective action $\Gamma_{k=0}$, after the integration of the 
ERGEs, which contains all
quantum contributions with an ultraviolet cutoff $\Lambda$. (In some cases 
it may be desirable to
work with a perturbatively improved action at $k = \Lambda$ \cite{19r}; 
this will not, however, be
employed here). \par

The boundary conditions concerning the dependence of $\Gamma_{k = \Lambda}$ 
on the auxiliary field
$H_{\mu \nu}^a$ are fixed by the way the source $K_{\mu \nu}^a$ is introduced 
in the path integral
(\ref{2.3}) \cite{18r}. The corresponding terms $\exp (K\cdot F + 
K \cdot K)$ can be rewritten with
the help of an auxiliary quantum field $H'$:

\beq
\label{2.13}
\exp \left ( K \cdot F + K \cdot K \right ) = {\cal N} \int {\cal D} H' 
\exp \left ( - {1 \over 4}
(H' - F)^2 + K \cdot H' \right ) \quad . \eeq 

\noi Now the auxiliary field appears on the same footing as the 
fundamental fields $A$, $c$ and
$\bar{c}$, with an action $S_{H'}$ given by $S_{H'} = {1 \over 4} 
(H' - F)^2$. (This way of
introducing the auxiliary field $H_{\mu \nu}^a$ is close to the field 
strength formulation of
Yang-Mills theories \cite{21r}.) It is actually convenient to rescale 
the auxiliary field
by a power of $\Lambda$, the only scale in the theory, in order to give 
it the appropriate dimension
of a bosonic field in $d= 4$. Then one finds for the boundary condition 
of the action $\Gamma_k$:

\beq
\label{2.14}
\Gamma_{k = \Lambda} = S_{YM} + S_g + \Delta S_k + {1 \over 4} 
(\Lambda H - F)^2 \quad ,
\eeq

\noi with $S_{YM}$, $S_g$ and $\Delta S_k$ as in eqs.~(\ref{2.4})--
(\ref{2.6}). For the parameters
$Z$, $n$, $m$, $h$, $\beta$ and $Z_g$ in (\ref{2.11}) this implies

\bea
\label{2.15}
&&Z(\Lambda) = 2 \ , \qquad n(\Lambda ) = \Lambda \ , \qquad m(\Lambda ) 
= \Lambda \quad , \nn \\
&&h(\Lambda ) = 0 \ , \qquad \beta (\Lambda ) = 0 \ , \qquad Z_g (\Lambda ) 
= 1 \quad .
\eea 

Note that, for arbitrary parameters in (\ref{2.11}), the field 
$H_{\mu \nu}^a$ could be eliminated
from (\ref{2.11}) by its equations of motion. This would lead to a 
dependence of $\Gamma_k$ on the
field strength $F_{\mu \nu}^a$ of the form

\beq
\label{2.16}
\Gamma_k (F) = \left ( Z_{eff}/4 \right ) \left ( F_{\mu \nu}^a \right )^2 
+ \dots \quad , 
\eeq
 
\noi where the dots denote terms of higher order in the covariant 
derivatives (induced by the terms
$\sim h , \beta$) and with

\beq
\label{2.17}
Z_{eff} = Z - {n^2 \over m^2} \quad .
\eeq

In terms of $Z_{eff}$ the boundary conditions (\ref{2.15}) become

\beq
\label{2.18}
Z_{eff}(\Lambda ) = 1
\eeq

\noi as they should. \par

Turning back to the case of general parameters in (\ref{2.11}), and 
leaving aside the gluon mass for
the moment, the terms shown in (\ref{2.11}) allow to obtain the full 
propagators for all fields. The
term proportional to $h$ and quadratic in $H_{\mu \nu}^a$ corresponds 
to a kinetic term for
$H_{\mu \nu}^a$. The expression $(\partial_{\mu} \widetilde{H}_{\mu 
\nu}^a)^2$ is
actually invariant under a gauge symmetry of the form $\delta H_{\mu 
\nu}^a = \partial_{[\mu}
\Lambda_{\nu ]}^a$ and would not be invertible, but the additional 
term proportional to $\beta$
serves as a ``gauge fixing term'' and allows for a finite $H_{\mu \nu}^a$ 
propagator (even for $m,n = 0$). \par

 The gauge fixing parameter $\alpha$ in (\ref{2.11}) can actually be 
chosen at will, and throughout
the paper we will work in the Landau gauge $\alpha = 0$, which is known 
to be a fixed point of the
ERGEs \cite{19r}.

Special attention in deriving the propagators has to be paid to the 
term ${n \over 2} F\cdot H$,
which induces a mixing between the gluon $A_{\mu}^a$ and the auxiliary 
field $H_{\mu \nu}^a$; this
modifies the gluon propagator considerably. Omitting the infrared cutoff 
in $\Delta S_k$ for
simplicity, one obtains

\bminiG{EDh}
\label{2.19a}
\left ( {\delta^2 \Gamma \over \delta \varphi_i(-p) \ \delta \varphi_j(p)} 
\right )^{-1}_{A_{\mu}^a,
A_{\nu}^b} &=& \delta_{ab} \left ( \delta_{\mu \nu} - {p_{\mu} p_{\nu} 
\over p^2} \right ) \cdot
P_A(p^2) \quad , \nn \\
P_A(p^2) &=& {p^2 \beta + m^2 \over p^2(Zm^2 - n^2) + Z \beta p^4} \quad 
, \eeeq 
\beeq
\label{2.19b}
\left ( {\delta^2 \Gamma \over \delta \varphi_i(-p) \ \delta \varphi_j(p)} 
\right )^{-1}_{A_{\mu}^a,
H_{\rho \sigma}^b} &=& - i \delta_{ab} \left ( p_{\rho} \ \delta_{\mu 
\sigma} - p_{\sigma} \
\delta_{\mu \rho } \right ) \cdot P_{AH}(p^2) \quad , \nn \\
P_{AH}(p^2) &=& {n \over p^2(Zm^2 - n^2) + Z \beta p^4} \quad ,
\eeeq
\beeq
\label{2.19c}
\left ( {\delta^2 \Gamma \over \delta \varphi_i(-p) \ \delta 
\varphi_j(p)} \right )^{-1}_{H_{\rho
\sigma}^a, H_{\kappa \lambda}^b} &=& \delta_{ab} \left ( \delta_{\rho 
\kappa} \ \delta_{\sigma
\lambda} - \delta_{\rho \lambda} \ \delta_{\sigma \kappa} \right ) 
P_{HH,1}(p^2) \nn \\  
+ \delta_{ab} \Big ( \delta_{\rho \kappa} \ p_{\sigma} \ p_{\lambda} 
- \delta_{\rho \lambda} \
p_{\sigma} \ p_{\kappa} &+& \delta_{\sigma \lambda} \ p_{\rho} \ 
p_{\kappa} - \delta_{\sigma
\kappa} \ p_{\rho} \ p_{\lambda} \Big ) P_{HH,2} (p^2) \quad , \nn \\
P_{HH,1}(p^2) &=& {1 \over hp^2 + m^2} \quad , \nn \\
P_{HH,2}(p^2) &=& {Z(h - \beta )p^2 + n^2 \over (hp^2 + m^2) (p^2 
(Zm^2 - n^2) + Z \beta p^4)} \ ,  
\eeeq 
\beeq 
\label{2.19d}
\left ( {\delta^2 \Gamma \over \delta \varphi_i (-p) \ \delta 
\varphi_j(p)} \right)^{-1}_{\bar{c}^a,
c^b} &=& \delta_{ab} \ P_g(p^2) \quad \nn \\
P_g(p^2) &=& {1 \over Z_g p^2} \quad . \emini

\noi From eqs. (2.19) and (\ref{2.15}) one easily obtains simple 
expressions for the
propagators at the starting point $k = \Lambda$. Let us now have 
a closer look at the gluon
propagator $P_A(p^2)$ in eq. (2.19a) (which would remain unchanged, 
if we would eliminate $H_{\mu
\nu}^a$ by its equation of motion). It depends on four parameters 
$Z$, $m$, $n$ and $\beta$ which,
in turn, depend on the scale $k$. Let us now \underbar{assume}, that 
at some scale $k_c$ (possibly
with $k_c = 0$) the parameters $Z$, $m$ and $n$ satisfy the relation

\beq
\label{2.20}
Z(k_c^2) - {n^2(k_c^2) \over m^2(k_c^2)} \equiv Z_{eff} (k_c^2) = 0 \quad .
\eeq

\noi Then the gluon propagator $P_A$ becomes

\beq
\label{2.21}
P_A(p^2) = {p^2 + \Lambda_c^2 \over Zp^4} \quad , \qquad \Lambda_c^2 = 
{m^2 \over \beta} \quad .
\eeq

\noi Thus we find that, for $p^2 \ll \Lambda_c^2$, the gluon propagator 
behaves like $1/p^4$.
Although the gluon propagator itself is a gauge dependent quantity, it is 
widely believed that such
a behaviour is a signal of confinement: G. West \cite{22r} has shown that 
a $1/p^4$ behaviour of
the gluon propagator in any gauge leads to an area law of the Wilson loop, 
and corresponding
results have been obtained in the context of Schwinger-Dyson equations 
for Yang-Mills theories
\cite{23r}. Subsequently we will adopt the conventional manner of 
speaking and call eqs.
(\ref{2.20}) and (\ref{2.21}) a confining behaviour. \par

Although we have not shown, at present, that such a confining behaviour 
actually appears for some value of $k_c$ (see section
3 below), we will proceed with the interpretation of its consequences. 
We wish to show that an
effective action $\Gamma_{k_c}$, which exhibits confinement in the 
sense of eqs. (\ref{2.20}) and
(\ref{2.21}), describes a physical situation in which monopole 
condensation has occurred. To this
end we wish to perform a duality transformation of the abelian 
subsector of $\Gamma_{k_c}$. (Note
that this does not imply that we neglect the non-abelian contributions 
to the ERGEs). For simplicity we will take
only one $U(1)$ subgroup into account, the generalization to several 
$U(1)$ subgroups being
straighfroward. \par

Neglecting $S_g$, $\Delta S_k$ and the terms not shown explicitely in 
(\ref{2.3}), and eliminating
$m^2$ by the relation (\ref{2.20}), the abelian projection of 
$\Gamma_{k_c}$ becomes

\beq
\label{2.22}
\Gamma_{k_c} = {1 \over 4Z} \left ( Z F_{\mu \nu} - n \ H_{\mu \nu} 
\right )^2 + {h \over 2} \left
( \partial_{\mu} \ \widetilde{H} \right )^2 + {\beta \over 2} \left 
( \partial_{\mu} \ H_{\mu \nu}
\right )^2 \ \ \ . \eeq

\noi Here $F_{\mu \nu}$ is the abelian field strength, and all gauge 
group indices have
disappeared. For the duality transformation we will also omit the 
``gauge fixing'' term $\sim
\beta$, since its presence would complicate the duality transformation 
considerably. \par

The equations of motion for the abelian gauge field $A_{\mu}$ and 
$H_{\mu \nu}$, respectively,
are then of the form

\[
\partial_{\mu} \left ( Z \ F_{\mu \nu} - n \ H_{\mu \nu} \right ) = 0 
\quad ,  
\]
\beq
\label{2.23}
h \ \varepsilon_{\mu \nu \rho \sigma} \ \partial_{\rho} \ 
\partial_{\lambda} \
\widetilde{H}_{\lambda \sigma} - F_{\mu \nu} + {n \over Z} H_{\mu \nu} = 
0 \quad . 
\eeq

\noi The Bianci identities corresponding to the fields $A_{\mu}$ and 
$H_{\mu \nu}$ are given by

\bea
\label{2.24}
\partial_{\mu} \ \widetilde{F}_{\mu \nu} &=& 0 \quad , \nn \\
\partial_{\mu} \ \partial_{\nu} \ \widetilde{H}_{\mu \nu} &= & 0 \quad .
\eea

\noi Now we introduce dual fields: the dual of $A_{\mu}$ is an abelian 
gauge field $B_{\mu}$ with
field strength $F_{\mu \nu}^B$, and the dual of the antisymmetric field 
$H_{\mu \nu}$ is a scalar
$\varphi$ (with a ``field strength'' $\partial_{\mu} \varphi$). The 
duality transformation mixes
the original fields $A_{\mu}/H_{\mu \nu}$ and the dual fields 
$B_{\mu}/\varphi$ in a non-trivial
way. It is given by  

\bea
\label{2.25}
\sqrt{Z} \ F_{\mu \nu}^B &=& {1 \over 2} \varepsilon_{\mu \nu \rho \sigma} 
\left ( Z \ F_{\rho
\sigma} - n \ H_{\rho \sigma} \right ) \quad , \nn \\
\partial_{\nu} \varphi - \widetilde{m} B_{\nu} &=& \sqrt{h} \ 
\partial_{\mu} \ \widetilde{H}_{\mu
\nu} \eea

\noi with $\widetilde{m} = n/2 \sqrt{Zh}$. \par

In terms of the dual fields the action $\widetilde{\Gamma}_{k_c}$ 
becomes

\beq
\label{2.26}
\widetilde{\Gamma}_{k_c}(B, \varphi ) = {1 \over 4} \left 
( F_{\mu \nu}^B \right )^2 + {1 \over 2} \left (
\partial_{\mu} \varphi - \widetilde{m} B_{\mu} \right )^2 \quad .
\eeq

\noi It generates the equations of motion for $B_{\mu}$ and $\varphi$, 
respectively,

\bea
\label{2.27}
&&{1 \over 2} \partial_{\mu} \ F_{\mu \nu}^B + \widetilde{m} \left 
( \partial_{\nu} \varphi -
\widetilde{m} B_{\nu} \right ) = 0 \quad , \nn \\
&&\sq \varphi - \widetilde{m} \ \partial_{\mu} \ B_{\mu} = 0 \quad .
\eea

\noi After inserting the duality transformations (\ref{2.25}) one 
finds that their content is equal
to the Bianci identities (\ref{2.24}). The Bianci identifies 
corresponding to the dual fields
$B_{\mu}$ and $\varphi$ are given by

\bea
\label{2.28}
\partial_{\mu} \ \widetilde{F}_{\mu \nu}^B &=& 0 \quad , \nn \\
\varepsilon_{\mu \nu \rho \sigma} \ \partial_{\rho} \ \partial_{\sigma} 
\varphi &=& 0 \quad .
\eea

\noi Again, after inserting the duality transformations (\ref{2.25}), 
one finds that their content
is equal to the original equations of motion (\ref{2.23}). \par

Let us now interpret the dual action $\widetilde{\Gamma}_{k_c}$, 
(\ref{2.26}): the scalar field
$\varphi$ couples like a Goldstone Boson to the dual (``magnetic'') 
gauge field $B_{\mu}$, i.e. we
are describing the Goldstone and gauge field sector of an abelian 
Higgs model, where the dual
``magnetic'' $U(1)$ is spontaneously broken. The dual action 
$\widetilde{\Gamma}_{k_c}$ can be
obtained from a full abelian Higgs model,

\beq
\label{2.29}
\widetilde{\Gamma}_{k_c} = {1 \over 4} \left ( F_{\mu \nu}^B \right )^2 
+ \left ( \partial_{\mu} +
ieB_{\mu} \right ) \phi^*\left ( \partial_{\mu} - ieB_{\mu} \right ) 
\phi + V(|\phi |^2)
\eeq

\noi in the limit where the vev of the complex scalar field $\phi$ 
is frozen,

\beq
\phi = {v \over \sqrt{2}} \ e^{i\varphi /v} \quad ,
\label{2.30}
\eeq

\noi and the identification $ev = \widetilde{m}$ is made. Since the 
charge $e$ of the complex field
$\phi$ (its ``electric'' coupling to the dual gauge field $B_{\mu}$) 
is to be interpreted as a
magnetic charge with respect to the original gauge field $A_{\mu}$, 
the spontaneous breakdown of
the dual $U(1)$ gauge symmetry due to dual electric charge condensation 
is to be interpreted as
magnetic charge condensation in the original theory (in spite of the 
fact that no non-trivial vacuum
appeared in the original theory~!). \par

Abelian Higgs models as duals of a confining low-energy effective 
actions of Yang-Mills theories
have been proposed before \cite{10r,11r}. Here, however, we have 
obtained the Goldstone degree of
freedom naturally as the dual of an antisymmetric field $H_{\mu \nu}$, 
which was introduced
originally for a quite different purpose, namely as an auxiliary 
field for the composite operator
$F_{\mu \nu}^a$. (During the preparation of this paper we became 
aware of a related proposal in
\cite{24r}). Also the idea of describing magnetically charged scalar 
fields as duals of an
antisymmetric field $H_{\mu \nu}$ is not new \cite{8r,9r}, but this 
approach was not applied to the
dual Meissner effect in \cite{8r,9r}. \par

In the following we emphasize that we do not have to assume the validity 
of the confining relation
(\ref{2.20}), which is required for the duality transformation to 
be possible, in an ad hoc
fashion: we have a dynamical scheme at our disposal, the Wilsonian 
ERGEs, which allows to obtain
$\Gamma_{k_c}$ - within certain approximations - from the bare 
Yang-Mills Lagrangian. This will be
the subject of the next section.

\mysection{The Wilsonian ERGEs}  
\hspace*{\parindent}
In this section we will discuss the Wilsonian ERGEs for the 
parameters ap\-pea\-ring (\ref{2.11}),
starting with the functional ERGE (\ref{2.10}) for the cutoff 
effective action. Since all
parameters $Z$, $n$, $m^2$, $h$, $\beta$ and $Z_g$ multiply 
expressions which contain quadratic
terms in the fields ($\sim AA$, $AH$, $HH$, or $\bar{c}c$) the 
corresponding ERGEs can be obtained
by expanding both sides of the functional ERGE (\ref{2.10}) to 
second order in the fields, and to
the appropriate zeroth, first or second order in the derivatives 
acting on the fields. From the
general structure of the ERGEs \cite{14r}-\cite{19r} one finds that 
the right-hand sides of the
ERGEs for parameters, which correspond to quadratic terms in the 
fields, involve only coefficients
of trilinear and quartic terms in the fields. Now we will define the 
approximation which we will
employ in the following: we will, on the r.h. sides of the ERGEs, 
take only those trilinear
couplings into account, which appear due to the non-abelian structures 
in the terms $(F_{\mu
\nu}^a)^2$, $F_{\mu \nu}H_{\mu \nu}$ and $\partial_{\mu}\bar{c} 
D_{\mu} c$ in (\ref{2.11}). These
depend on $Z$, $n$, $Z_g$ and the coupling $\bar{g}$ (in 
eq.~(\ref{2.12}) and in $F_{\mu \nu}^a$),
but the Slavnov-Taylor identities fix, in this approximation 
and in the Landau gauge \cite{19r}
$\bar{g}$ to be

\beq
 \label{3.1}
\bar{g} = g_0/Z_g
\eeq  

\noi with $g_0$ (which is equal to the ``bare'' coupling in 
$\Gamma_{k=\Lambda}$) independent of
$k$. If one eliminates $H_{\mu \nu}^a$ by its equation of motion, 
and with an appropriate field
redefinition of $A_{\mu}^a$, one easily finds that a physical 
running gauge coupling $g_{phys}$
(defined at the 3-gluon, 4-gluon or ghost-gluon vertex) is given by

\beq
g_{phys} = \bar{g}/\sqrt{Z_{eff}} = g_0/\left ( Z_g \cdot 
\sqrt{Z_{eff}} \right )
\label{3.2}
\eeq

\noi with $Z_{eff}$ as in eq. (\ref{2.17}). \par

The approximation consists, in particular, in neglecting the 
contributions from the vertices
appearing in the terms $(D_{\mu}\widetilde{H}_{\mu \nu})^2$ and 
$(D_{\mu}H_{\mu \nu})^2$ (remember,
however, that $h = \beta = 0$ at the starting point) and the 
contributions from terms not shown
explicitely in (\ref{2.11}). The presence of some such terms 
in $\Gamma_k$ is actually required by
the modified Slanov-Taylor identities \cite{15r,16r}. (Amongst 
others, the presence of a gluon mass
term is necessary. The modified Slavnov-Taylor identities imply, 
however, that this term vanishes
in the limit $k^2 \to 0$). Note that an approximation of the 
ERGEs of the present type
does not imply that $\Gamma_k$ violates these identities; we 
rather neglect contributions on the
right-hand side of the ERGEs of parameters, which themselves are 
\underbar{not} constrained by the
Slavnov-Taylor identities. \par

With $\bar{g}$ determined by (\ref{3.1}) we have thus obtained a 
closed system of ERGEs for the 6
parameters $Z$, $n$, $m^2$, $h$, $\beta$ and $Z_g$. This is the 
minimal set which allows a) to
parametrize the bare action (or the bare Lagrangian), and to obtain 
the correct 1-loop $\beta$
functions, and b) to parametrize a ``confining'' action where the 
parameters satisfy
eq.~(\ref{2.20}). Since actually the quartic vertex without 
derivatives in $(F_{\mu \nu}^a)^2$ does
not contribute to the ERGEs in this approximation, all 6 ERGEs 
are of the same diagrammatic form as
shown in fig.~1. \par

The type of external fields $\{ \varphi_a, \varphi_b \}$ and the 
powers of derivatives
$\partial_{\mu}$ acting on them depend on the parameter, whose ERGE 
one is considering: \par

 -- $\{ \varphi_a , \varphi_b \} = \{ A, A\}$ and ${\cal O}\left 
( \partial^2_{\mu} \right )$ in the
case of $Z$, \par

-- $\{ \varphi_a , \varphi_b \} = \{ A, H\}$ and ${\cal O}\left 
( \partial_{\mu} \right )$ in the
case of $n$, \par

-- $\{ \varphi_a , \varphi_b \} = \{ H, H\}$ and ${\cal O}\left 
( \partial_{\mu}^0 \right )$ in the
case of $m^2$, \par

-- $\{ \varphi_a , \varphi_b \} = \{ H, H\}$ and ${\cal O}\left 
( \partial^2_{\mu} \right )$ in the
cases of $h$ and $\beta$, \par

-- $\{ \varphi_a , \varphi_b \} = \{ \bar{c}, c\}$ and ${\cal O}\left 
( \partial^2_{\mu} \right )$ in
the case of $Z_g$. \par

\noi Note that two different Lorentz index structures are possible in 
the case $\{H, H\}$ and ${\cal
O}\left ( \partial^2_{\mu} \right )$, which have to be distinguished 
in order to separate the
contributions to $h$ and $\beta$. The internal fields $\{\varphi_i , 
\varphi_j\}$  attached to the insertion of $\partial_k R$ in fig. 1 
have necessarily to
be of the type $\{ A, A\}$ or $\{ \bar{c} , c\}$, because only for 
these fields infrared cutoff terms
$\Delta S_k$ exist. We have to take into account, however, mixed 
propagators of the type $\{\varphi_e
, \varphi_i \} = \{H, A\}$ or $\{ \varphi_c , \varphi_d \} = \{H, A \}$ 
etc., which lead to many
different contributions even if the external fields $\{\varphi_i, 
\varphi_j \}$ are fixed. \par

In order to exhibit the resulting ERGEs we employ the following notations: 
$P_A$, $P_{AH}$ etc.
denote the propagator functions shown in eqs. (2.19), with the infrared 
cutoffs

\beq
\label{3.3}
R^k (p^2) = {e^{-p^2/k^2} \over 1 - e^{-p^2/k^2}}
\eeq

\noi due to $\Delta S_k$ and the form of the functions $R_{\mu \nu}^k(p^2)$ 
and $R_g^k(p^2)$ (cf.
eq. (\ref{2.7})) restored. This is easily done by replacing $Z \to Z + R^k$ 
and $Z_g \to Z_g +
R^k$ in eqs. (2.19). Actually the exhibited form of the equations is 
general and allows for
different choices of the cutoff function (\ref{3.3}). (Our choice 
guarantees that all integrals
appearing on the r.h. sides of the ERGEs are ultraviolet finite; 
however, we were not able to
perform these momentum integrations analytically.) \par

Because of the required development in powers of derivatives (or 
external momenta) the quantities
$p^2 {\partial P_A \over \partial p^2}$ etc. appear frequently, 
and we define for convenience

\beq
\label{3.4}
P' \equiv p^2 \ {\partial P \over \partial p^2} \quad , \qquad P'' 
\equiv p^4 \ {\partial^2 P
\over (\partial p^2)^2}
\eeq

\noi in all cases $P_A$, $P_{AH}$ etc. Finally we define $\partial_k 
\equiv k^2 \partial /
\partial k^2$, and performed the combinatories of the gauge group 
indices in the case of a
$SU(N)$ gauge group. The 6 ERGEs then become  

\bminiG{EDh}
\label{3.5a}
\partial_k \ Z &=& {N\bar{g}^2 \over 16 \pi^2} \int_0^{\infty} dp^2 
\ p^4 \ \partial_k \ R^k \Big
[ Z^2 \left ( {31 \over 6} P_A^3 + 3P_A^2 P'_A + P_A^2 P''_A \right ) 
\nn \\
&&- n Z \Big ( {16 \over 3} P_A^2 \ P_{AH} + {8 \over 3} P_A \ P_{AH} \
 P'_A + P_A \ P_{AH} \
P''_A \nn \\
&& + {10 \over 3} P_A^2 \ P'_{AH} + P_A^2 \ P''_{AH} \Big ) \nn \\
&&+ n^2 \Big ( P_A \ P_{AH} \left ( {3 \over 2} P'_{AH} + {1 \over 2} 
P''_{AH} \right ) \Big ) \nn
\\
&&+ P^2_{AH} \left ( {5 \over 12} P_A + {7 \over 12} P'_A + {1 \over 4} 
P''_A \right ) \nn \\
&&+ P_A^2 \left ( {9 \over 4} P'_{HH,1} + {13 \over 12} P''_{HH,1} + 
{3 \over 4} P_{HH,2} + {11
\over 12} P'_{HH,2} + {1 \over 4} P''_{HH,2} \right ) \nn \\
&& + {1 \over 6} Z_g^2 \ P_g^3 \Big ] \quad , \eeeq \beeq \label{3.5b} 
\partial_k \ n =
{N \bar{g}^2 \over 16 \pi^2} \int_0^{\infty} dp^2 \ p^4 \ \partial_k 
\ R^k \ n \ P_A^2 \left ( {5
\over 2} Z \ P_A - {3 \over 2} n\ P_{AH} \right ) \quad ,  \eeeq \beeq
\label{3.5c}
\partial_k \ m^2 = {N \bar{g}^2 \over 16 \pi^2} \int_0^{\infty} dp^2 \ 
p^4 \ \partial_k \ R^k \ n^2
\ P_A^3 \quad , \eeeq
\beeq
\label{3.5d}
\partial_k \ Z_g = {N\bar{g}^2 \over 16 \pi^2} \int_0^{\infty} dp^2 \ p^4 
\ \partial_k \ R^k \
Z_g^2 \ P_A \ P_g {3 \over 4} \left ( P_A + P_g \right ) \quad , \eeeq
\beeq
\label{3.5e}
\partial_k \ h &=& {N \bar{g}^2 \over 16 \pi^2} \int_0^{\infty} dp^2 \ p^2 
\ \partial_k \ R^k \ n^2 \
P_A^2 \left ( {1 \over 6} P_A + {4 \over 3} P'_A + {2 \over 3} P''_A 
\right ) \quad , \nn \\ 
\partial_k \beta &=& {1 \over 2} \partial_k h \quad . \emini

\noi Note that all integrals are trivially UV finite since, with the 
present choice of $R^k$,
$\partial_kR^k$ decreases exponentially for large $p^2$, and IR 
finiteness is ensured by the
presence of the IR cutoff terms in the propagators. \par

Let us first insert the ``starting point values'' (\ref{2.1}) for 
the pa\-ra\-me\-ters on
the r.h. sides of the ERGEs (3.5). Then all integrals can be 
performed analytically, and we
should obtain the correct 1-loop $\beta$ functions. We find

\bea
\label{3.6}
&&\partial_k \ Z = {N g_0^2 \over 16 \pi^2} \cdot {31 \over 6} \quad , 
\nn \\
&&\partial_k \ n = {N g_0^2 \over 16 \pi^2} \cdot {7 \over 4} \Lambda 
\quad , \nn \\
&&\partial_k \ m^2 = {N g_0^2 \over 16 \pi^2} \cdot {1 \over 2} 
\Lambda^2 \quad , \nn \\
&&\partial_k \ h = - {N g_0^2 \over 16 \pi^2} \cdot {7 \over 12}  
\quad , \nn \\
&&\partial_k \ Z_g = {N g_0^2 \over 16 \pi^2} \cdot {3 \over 4}  \quad . 
\eea

\noi This leads, indeed, to

\beq
\label{3.7}
\partial_k \ Z_{eff} = \partial_k \left ( Z - {n^2 \over m^2} \right ) = 
{N g_0^2 \over 16 \pi^2}
\cdot {13 \over 6} \quad ,  \eeq

\noi thus we obtain the correct 1-loop $\beta$ function for $g_{phys}$ 
as defined in eq.
(\ref{3.2}). From eqs. (\ref{3.6}) and (\ref{3.7}) one concludes that 
$Z$, $n$, $m^2$, $Z_g$ and
$Z_{eff}$ start to decrease with decreasing $k^2$, whereas $h$ (and 
hence $\beta$, cf.
(\ref{3.5e})) start to increase. \par

The complete system of ERGEs (3.5) can be integrated numerically, 
with the choice of $g_0
\equiv g_{phys}(\Lambda^2)$ as the only freedom at the starting point. 
In fig.~2 we show a
typical result obtained with $g_0 = 1.3$. We plot directly $Z_{eff}(k^2) 
= Z(k^2) -
n^2(k^2)/m^2(k^2)$ and, for convenience, $\alpha_{phys}(k^2) = 
g_0^2/(4\pi Z_g^2
(k^2)Z_{eff}(k^2))$ versus $-\ell n (k^2/\Lambda^2)$. We find that 
at a small, but finite value
of $k^2 = k_c^2$ ($=1,94\cdot 10^{-5} \Lambda^2$ in the present case) 
both $Z_g(k_c^2)$ and
$Z_{eff}(k_c^2)$ vanish, thus $\alpha_{phys}(k^2)$ runs into a Landau 
singularity and the r.h.
sides of the ERGEs (3.5) explode. The parameters $Z(k^2)$, $n(k^2)$ and 
$m^2(k^2)$, however, remain finite and non-vanishing for $k^2 \to k_c^2$ 
(in the present
case we have $Z(k_c^2) \cong 0.24$, $n(k_c^2) \cong .43 \cdot \Lambda$, 
$m^2(k_c^2) \cong 0,76\cdot
\Lambda^2$). \par

 From the previous section, in particular from eq.~(2.20) and the 
subsequent discussion, we know
that a vanishing of $Z_{eff}(k_c^2)$ corresponds to a confining form 
of the gluon propagator. In
the corresponding expression eq.~(\ref{2.21}) a dimensionful quantity 
$\Lambda_c^2 = m^2/\beta$
appears which could be related to the slope of a linearly rising 
$q\bar{q}$-potential
obtained by a ``dressed'' one-gluon exchange (with a gluon propagator 
as in eq.~(\ref{2.21})) in
the non-relativistic limit. As such it should be independent from the 
starting scale $\Lambda$.
However, $\Lambda$ is the only dimensionful scale in our approach, hence 
all dimensionful
quantities are necessarily proportional to $\Lambda$. The physical 
meaning of $\Lambda$ is
implicit in the choice of the bare coupling $g_0$: $g_0 = 
g_{phys}(\Lambda^2)$. Given the
$\Lambda$-loop $\beta$ function for $g_{phys}$ (the 2-loop 
$\beta$ function is not obtained
correctly within our approximation), the combined dependence 
of $\Lambda_c^2$ on $\Lambda^2$
and $g_0^2$ should thus be of the form 

\beq
\label{3.8}
\Lambda_c^2 = \Lambda^2 \ e^{-(16 \pi^2/11g_0^2)} \left ( 1 + {\cal O} 
\left ( {g_0^2 \over 4
\pi } \right ) \right ) \quad .
\eeq

We have calculated the quantity

\beq
\label{3.9}
C = {\Lambda^2 \over \Lambda_c^2 (k_c^2)} \ e^{-(16 \pi^2/11g_0^2)} = 
{\Lambda^2 \beta
(k_c^2) \over m^2 (k_c^2)} \ e^{-(16 \pi^2/11g_0^2)}
\eeq

\noi for various choices of $g_0$, where the pattern of ERGE flow for 
all parameters is
always of the form presented in the case $g_0 = 1.3$ (but with 
different results for
$k_c^2/\Lambda^2$, of course): $g_0 = 1.3$: $C = 1.68 \cdot 10^{-2}$; 
$g_0 = 1.4$: $C =
1.38 \cdot 10^{-2}$; $g_0 = 1.5$: $C = 1.20\cdot 10^{-2}$; $g_0 = 1.6$: 
$C = 1,09 \cdot
10^{-2}$; $g_0 = 1.7$: $C = 1.01\cdot 10^{-2}$. Thus, although the 
corresponding values of
$\Lambda_c^2(k_c^2)/\Lambda^2$ vary over 2 orders of magnitude, $C$ 
remains approximately
constant with an accuracy in agreement with eq.~(\ref{3.8}). (For 
even larger values of
$g_0$ the ratio $\Lambda_c^2/\Lambda^2$ becomes larger than 1, i.e. 
the UV cutoff $\Lambda$
is below the ``confinement scale'' $\Lambda_c$, whereas for smaller 
values of $g_0$ the
number of discrete steps for the integration of the ERGEs ($\sim 
\Lambda^2/k_c^2$) becomes
extremely large without any new insight being gained). \par

Actually the fact that the ``confining'' relation $Z_{eff}(k_c^2) = 0$ 
and hence a Landau
singularity in $\alpha_{phys}(k_c^2)$ appears already at a finite 
value of $k^2 = k_c^2$ is
to be considered as an artefact of the present approximation to the 
ERGEs, which does
evidently not yet capture the entire infrared dynamics of Yang-Mills 
theories (as, e.g., a
possible gluon condensate). Focusing on the three parameters $Z$, 
$n$ and $m^2$, which
appear in the ``confining'' relation (\ref{2.20}), it is very 
interesting, however, that an
analytic result can be obtained, which is independent of the other 
parameters: let us
consider the r.h. sides of the corresponding ERGEs in the case where 
the relation $Z(k^2) -
n^2(k^2)/m^2(k^2) = 0$ is fulfilled. In addition we assume that the 
scale $k^2$ is already
far below the ``confinement scale'' $\Lambda_c^2 = m^2/\beta$, $k^2 
\ll \Lambda_c^2$.
Since only momenta with $p^2 \lsim k^2$ contribute to the integrals 
of the ERGEs
(3.5) (remember the exponential damping), this assumption allows to 
simplify the
propagators considerably and to obtain analytic results. We find         

\bea
\label{3.10}
&&\partial_k \ Z = {N\bar{g}^2 \over 16 \pi^2} \cdot {\Lambda_c^2 
\over k^2} \cdot {Z \over 4}
\quad , \nn \\
&&\partial_k \ n = {N\bar{g}^2 \over 16 \pi^2} \cdot {\Lambda_c^2 
\over k^2} \cdot {n \over 4}
\quad , \nn \\
&&\partial_k \ m^2 = {N\bar{g}^2 \over 16 \pi^2} \cdot {\Lambda_c^2 
\over k^2} \cdot {m^2 \over 4}
\quad . 
\eea

One easily derives

\beq
\label{3.11}
\partial_k \ Z_{eff} = \partial_k \left ( Z - {n^2 \over m^2} \right ) 
= {N\bar{g}^2 \over 16
\pi^2} \cdot {\Lambda_c^2 \over k^2} \cdot {Z_{eff} \over 4} \quad ,
 \eeq

\noi i.e. whereas the individual parameters $Z$, $n$, and $m^2$ continue 
to run in a way which
depends on the running of $\bar{g}^2$, $Z_{eff} = 0$ is a quasi-fixed 
point of the ERGEs. Moreover,
in agreement with our numerical analysis, this quasi-fixed point is 
infrared attractive. Were it not
for the singular behaviour of $\bar{g}^2$ for small $k^2$ due to the 
vanishing of $Z_g$, the
integrated eq.~(\ref{3.11}) would lead to a smooth vanishing of 
$Z_{eff}$ for $k^2 \to 0$. Again,
however, we cannot check in how far the quasi-fixed point $Z_{eff} = 0$ 
depends on the present
approximations to the ERGEs.

\mysection{Conclusions}  
\hspace*{\parindent} 
The intentions of the present paper are twofold: first, we proposed an 
effective low energy action
for Yang-Mills theories, which involves an additional an\-ti\-sym\-me\-tric 
tensor field $H_{\mu
\nu}^a$ introduced as an auxiliary field for the composite local operator 
$F_{\mu \nu}^a$. We have
seen that, for a particular relation between 3 of the 6 parameters in this 
action, a confining gluon
propagator in the sense of a $1/p^4$ behaviour for $p^2 \to 0$ is obtained. 
On the other hand, we
were able to relate this behaviour to a condensate of monopoles: the 
presence of the field $H_{\mu
\nu}^a$ in the original action generates automatically the presence of 
a Goldstone boson $\varphi$
in the dual action (of the abelian subsystem), which is thus of the form 
of an abelian Higgs model
in the broken phase (with frozen radial excitations). This ``dual'' 
role of the auxiliary field
$H_{\mu \nu}^a$ is certainly a particular feature of gauge theories 
in $d = 4$ dimensions. An
amazing and conceptually important feature is the fact that, from 
the point of view of the original
action, the confining behaviour of the gluon propagator is 
\underbar{not} directly related to a
non-trivial vacuum (it can be speculated, however, that a 
vanishing of $Z_{eff}$ is a necessary
condition for the formation of a gluon condensate). \par

Second, we have discussed a dynamical scheme, which allows to 
compute the corresponding low energy
effective action from the bare Yang-Mills Lagrangian: the Wilsonian 
ERGEs including auxiliary
fields. Since this constitutes, a priori, an exact formalism, it 
allows for various approximations
in order to turn it into a closed system of differential equations. 
These approximations can be
chosen such that they correspond to the physical problem under 
consideration. Here we did not solve
the ERGEs for the parameters corresponding to higher dimensional 
or non-BRST-invariant terms in the
effective action, whose knowledge is not required for the check of the 
``confining relation''
eq.~(\ref{2.20}). The approximation consisted then in the neglect of 
the contributions of these
parameters to the ERGEs of the 6 important parameters 
$Z$, $n$, $m^2$, $h$, $\beta$ and $Z_g$. We
have seen that, within this approximation, the relation (\ref{2.20}) 
constitutes indeed an
infrared attractive quasi-fixed point of the ERGEs. On the other hand, 
the fact that the relation
(\ref{2.20}) is assumed already at a finite scale $k^2 = k_c^2$, and 
the physical gauge coupling
runs into a Landau singularity, is obviously an artifact of the 
approximation. We have certainly
not yet solved the problem of finding approximations to the system 
of ERGEs, which allow to
obtain numerically reliable results in the infrared. \par

However, the present approach may point into a direction where 
this problem may be solvable. We
have seen that the duality transformation of the abelian subsystem 
provides us with the action of
an abelian Higgs model in the broken phase. This model has trivial 
infrared fixed points in the
Wilsonian sense, since it contains no massless physical fields which 
could contribute to the ERGE
flow in the limit $k^2 \sim p^2 \to 0$. Duality transformations can 
also be considered for full
non-abelian actions, see \cite{25r} for first steps in this direction. 
Here the genuine problem is
that one obtains actions, which are non-local, non-polynomial in the 
fields, and/or involve
additional auxiliary degrees of freedom. On the other hand, all these 
features are already present
in Wilsonian effective actions in general, and the Wilsonian functional 
ERGE is capable to cope
with then. Let us now speculate that the result of a duality 
transformation of a full non-abelian
action, again in the presence of auxiliary fields $H_{\mu \nu}^a$, 
leads us again to dual actions
which contain no massless physical fields.  Then we would obtain 
automatically infrared fixed
points for the Wilsonian ERGEs for the \underbar{dual} action, 
which allow to obtain numerically
reliable results in the infrared. It then remains to reinterpret 
the infrared effective action of
the dual theory in terms of the original theory in order to extract 
its physical consequences.
Although we are aware of the speculative nature of this approach 
it is presently under
investigation. \\

\section*{Acknowledgements} 
\hspace*{\parindent} We have profited from numerous discussions 
with B. Bergerhoff, D. Jungnickel, A. Weber and C.
Wetterich.

\vspace{5cm}
\begin{figure} [h]
$$\epsfbox{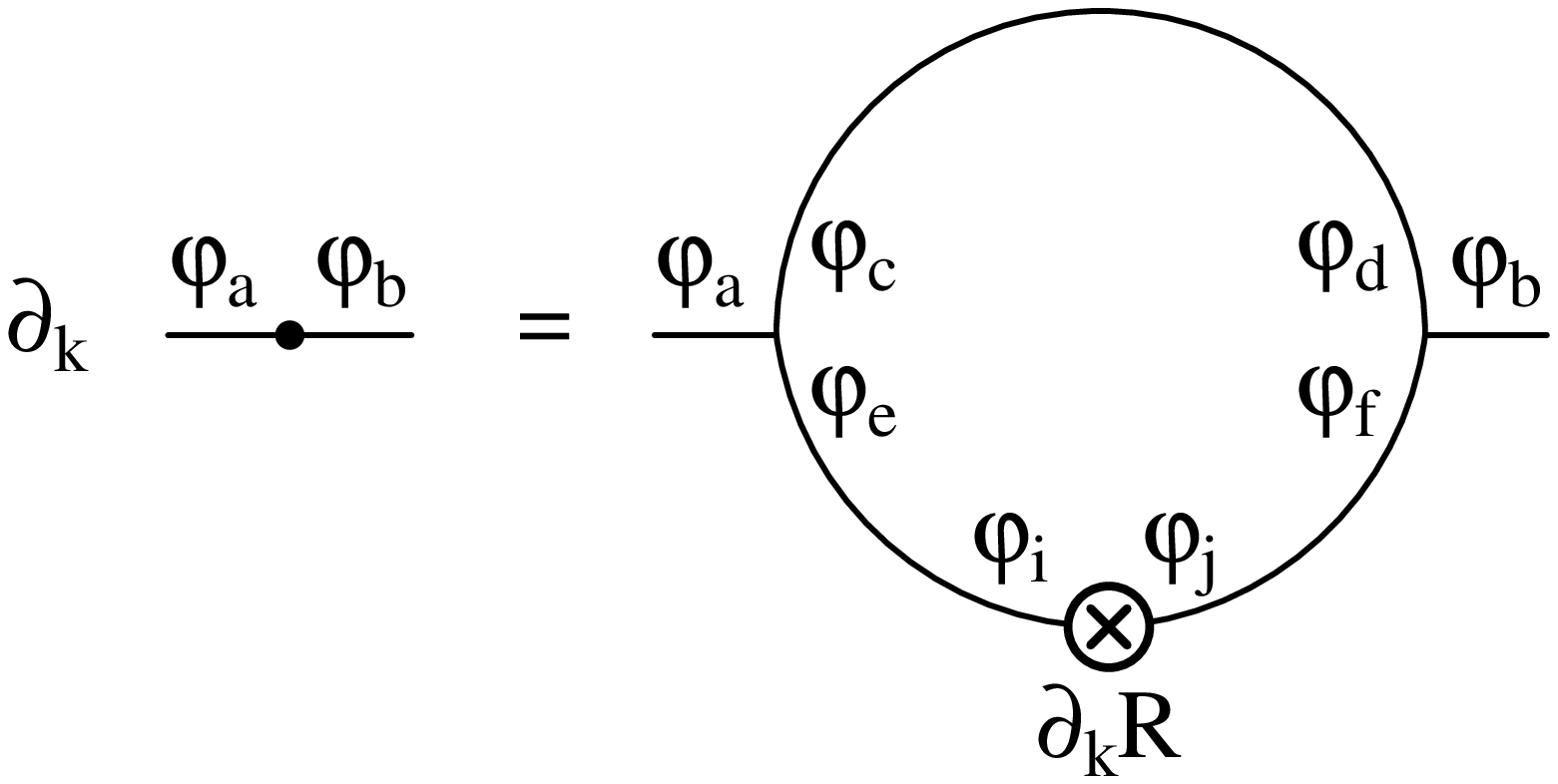}$$
\caption{ 
Diagrammatic form of the ERGEs for the parameters
corresponding to terms quadratic in the fields in $\Gamma_k$. 
The external fields $\varphi_a$,
$\varphi_b$ have to be chosen to be $A$, $H$, $c$, or $\bar{c}$ 
as described in the text. The
crossed circle denotes an insertion of the function $\partial_k 
R^k(p^2)$. 
}

\end{figure}

\begin{figure} [h]
$$\epsfbox{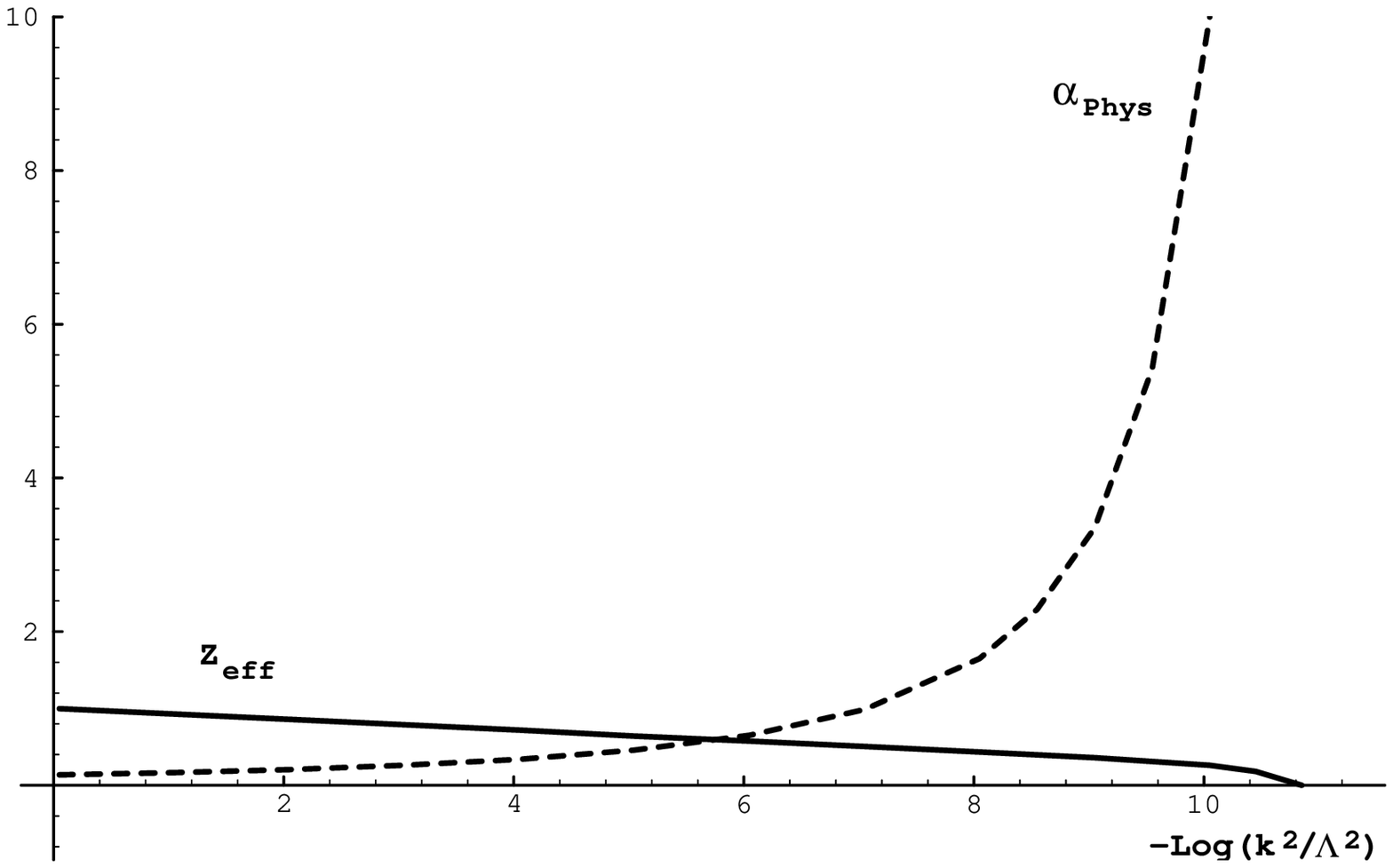}$$
\caption{
Result of the numerical integration of the ERGEs (3.5) with 
$g_0 = 1.3$. 
$Z_{eff}(k^2)$ and $\alpha_{phys}(k^2)$ are defined in 
eq.~(\protect{\ref{2.17}}) and below eq.~(\protect{\ref{3.7}}), 
respectively. The ERGE flow starts at 
$- \ell n (k^2/\Lambda^2) = 0$ and stops due to the Landau 
singularity at 
$- \ell n (k^2_c/\Lambda^2) \sim 10.85$. 
}

\end{figure}

\newpage
\begin{thebibliography}{99} \par \vskip 5 truemm 
\bibitem{1r} S. Mandelstam, Phys. Rep. {\bf 23C} (1976) 245; 
G. t'Hooft, Nucl. Phys. {\bf B190}
(1981) 455. 
  
\bibitem{2r} Lattice 96, Proceedings; Nucl. Phys. B (Proc. Suppl.) 
{\bf 53} (1997). 
  
\bibitem{3r} A. Polyakov, JETP Lett. {\bf 20} (1974) 194; \\
 G. t'Hooft, Nucl. Phys. {\bf
B79} (1974) 276. 

  \bibitem{4r} M. Blagojevic, P. Senjanovic, Phys. Rep. {\bf 157} 
(1988) 233. 
 
\bibitem{5r} A. Di Giacomo, M. Mathur, Phys. Lett. {\bf B400} 
(1997) 129. 
  
\bibitem{6r} K. Bardakci, S. Samuel, Phys. Rev. {\bf D18} (1978) 
2849; \\
E. Akhmedov, M. Chernodub, M. Polikarpov, M. Zubkov, Phys. Rev. 
{\bf D53} (1996) 2087. 
 
\bibitem{7r} D. Zwanziger, Phys. Rev. {\bf D3} (1971) 880. 
 
 \bibitem{8r} M. Mathur, H. Sharatchandra, Phys. Rev. Lett. {\bf 66} 
(1991) 3097. 

  \bibitem{9r} K. Lee, Phys. Rev. {\bf D48} (1993) 2493. 
  
\bibitem{10r} S. Maedan, Y. Matsubara, T. Suzuki, Prog. Theor. Phys. 
{\bf 84} (1990) 130 and refs. therein.  
 
\bibitem{11r} M. Baker, J. Ball, F. Zachariasen, Phys. Rep. {\bf 209} 
(1991) 73, Int. J. Mod. Phys.
{\bf A11} (1996) 343; \\
M. Baker, J. Ball, N. Brambilla, G. Prosperi, F. Zachariasen, 
Phys. Rev. {\bf D54} (1996) 2829.

\bibitem{12r} K. Wilson and I. Kogut, Phys. Rep. {\bf 12} (1974) 75; \\
F. Wegner, in: Phase Transitions and Critical Phenomena, Vol. 6, 
eds. C. Domb and M. Green
(Academic Press, NY 1975).

\bibitem{13r} S. Weinberg, in: Proceedings of the 1976 International 
School of Subnuclear Physics,
Erice, ed. A. Zichichi; \\
J. Polchinski, Nucl. Phys. {\bf B231} (1984) 269.

\bibitem{14r} G. Keller, C. Kopper, Phys. Lett. {\bf B273} 
(1991) 323; \\
G. Keller, C. Kopper, M. Salmhofer, Helv. Phys. Acta {\bf 65} 
(1997) 32; \\
C. Wetterich, Phys. Lett. {\bf B301} (1993) 90; \\
C. Wetterich, M. Reuter, Phys. Lett. {\bf B334} (1994) 412; \\
T. Morris, Int. J. Mod. Phys. {\bf A9} (1994) 2411; \\
M. D'Attanasio, T. Morris, Phys. Lett. {\bf B378} (1996) 213; \\
For more references see J. Comellas, Y. Kubyshin, E. Moreno, 
Nucl. Phys. {\bf B490} (1997) 653.
 
\bibitem{15r} C. Becchi, in: Lectures given at the Summer School 
in Theoretical Physics, Parma
1991-1993, eds. M. Bonini, G. Marchesini and E. Onofri, 
Parma University 1993; \\
M. Bonini, M. D'Attanasio, G. Marchesini, Nucl. Phys. {\bf B418} 
(1994) 81, Nucl. Phys. {\bf B421}
(1994) 429, Nucl. Phys. {\bf B437} (1995) 163, Phys. Lett. 
{\bf B346} (1995) 87. 
   
\bibitem{16r} U. Ellwanger, Phys. Lett. {\bf B335} (1994) 364. 

\bibitem{17r} U. Ellwanger, Z. Phys. {\bf C58} (1993) 619. 

\bibitem{18r} U. Ellwanger, C. Wetterich, Nucl. Phys. {\bf B423} 
(1994) 137. 

\bibitem{19r} U. Ellwanger, M. Hirsch, A. Weber, Z. Phys. {\bf C69} 
(1996) 687 and hep-ph/9606468,
to appear in Z. Phys. {\bf C}; \\
U. Ellwanger, hep-ph/9702309, to appear in Z. Phys. {\bf C}; \\
B. Bergerhoff, C. Wetterich, hep-ph/9708425. 

\bibitem{20r} R. Haymaker, Riv. Nuovo Cim. {\bf 14} (1991), No.~8.

\bibitem{21r}M. Holperu, Phys. Rev. {\bf D16} (1977) 1798; \\
M. Quandt, H. Reinhardt, hep-th/9707185 and refs. therein.

\bibitem{22r} G. West, Phys. Lett. {\bf B115} (1982) 468.

\bibitem{23r} C. Roberts and A. Williams, Prog. in Part. and Nucl. 
Physics {\bf 33} (1994) 477, and references therein.

\bibitem{24r} K. Kondo, preprint CHIBA-EP-99, hep-th/9709109.

\bibitem{25r} K. Itabashi, Prog. Theor. Phys. {\bf 65} (1981) 1423; \\
L. Mizrachi, Annals Phys. {\bf 153} (1984) 417 and Phys. Lett. {\bf B139} 
(1984) 374; \\
A. Caticha, Phys. Rev. {\bf D37} (1988) 2323. 

\end{thebibliography}
\end{document}